# Reconfigurable high Gain split Ring Resonator Microstrip Patch Antenna

**Japit S. Sonagara\*, Karan H. Shah[¥], Jaydeep D. Suvariya and Shobhit K. Patel**
Marwadi Education Foundation Group of Institutions, Rajkot, India.
*japit.sonagara13849@ marwadieducation.edu.in
[¥]karan.shah13312@marwadieducation.edu.in

**ABSTRACT**: In this paper, reconfigurable high gain split ring resonator microstrip patch antenna is designed and analysed. The aim to design such type antenna is to achieve multiband application which is the demand of current technology in frequency reconfiguration within single antenna. Here microstrip patch antenna with rectangle shape of patch with patch dimension 11.6×11.6 mm$^2$ is analysed. The proposed design is tuned with two bands in the frequency range of 5-9 GHz depending on the geometric specification of antenna and the location of feed which can be used for multiband applications. Design results of VSWR, return loss ($S_{11}$), bandwidth and gain are shown in this paper which is obtained by high frequency structure simulator (HFSS) which is used for simulating microwave passive components.

## I. INTRODUCTION

With rapid increasing technology in the field of recent wireless communication system, the designing of antenna with multiple frequencies and decreasing size is required [1]. It is necessary to design multiple frequencies antenna to complete the need of different application in current scenario [2]. This is done by microstrip patch antenna; it is attractive due to many advantages like small size, less weight, low cost, low profile, easy production and easy addition of feed network. Microstrip antennas provide more numbers of physical parameters as compared to conventional microwave antennas. Microstrip antenna also satisfies the need of WLAN. Microstrip patch antenna is sub divided into four basic categories as shown in figure1.

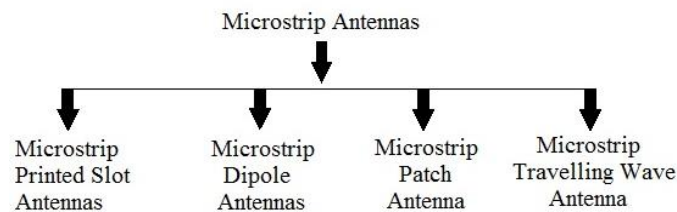

Fig. 1. Types of Microstrip Antenna

One of the main functions of antenna is convert electrical signal to electromagnetic waves and vice versa. Feed of an antenna is very important. In Microstrip antenna following feeding techniques are widely used, coaxial probe Feed, microstrip line, aperture coupling and proximity coupling in our case we have used a coaxial feeding method. It contains two conductors, out of which inner conductor of the coaxial is attached with the patch while the other one is attached with the ground plane. Advantages of coaxial feeding are easy fabrication, easy to match, low counterfeit radiation and its disadvantage is narrow bandwidth, difficult to replica especially for thick substrate. Main disadvantage of conventional microstrip antenna are narrow bandwidth and in current scenario of wireless communication it requires large bandwidth, so bandwidth enhancement is most important in present day. This drawback of microstrip antennas can be overcome by introducing different slots in patch. It also gives the benefits of size reduction [3-11]. With introducing slots in microstrip patch antenna it provides following advantage like, bandwidth of microstrip patch antenna is increased, efficiency is increased, improved VSWR, reduced size and increased gain [12-13]. It also gives output at multiple beam so it is helpful for multiple frequency application e.g. reconfigurable antenna. There are various types of slots available like C slot patch, H slot patch, S slot patch, L slot patch, U slot patch, V slot patch E slot patch etc. [14]. In our case we used C slot patch with complementary split ring resonator microstrip patch antenna. Design of complementary split ring resonator microstrip patch antenna is shown in a figure2.

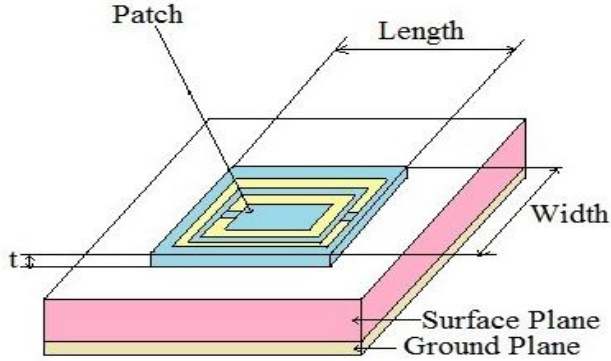

Fig. 2. Structure of Microstrip Patch Antenna.

## II. Design and Modelling

This section gives the introduction of our antenna design. Here the first step is to design the length and width of patch. In our design split ring resonator is taken on the patch. Top view and side view of proposed split ring resonator is shown in a fig. 3 and fig. 4 respectively. The width of the patch is designed using below equation. Here $f_r$ is the center frequency, r is the relative permittivity and c is speed of light. Using below equations the length of patch can be calculated. Here h is the height of substrate.

Following equations are used to design the length and width of the patch

$$w = \frac{c}{2f_r}\sqrt{\frac{2}{\epsilon_r+1}} \tag{1}$$

$$\epsilon_{r_{eff}} = \frac{\epsilon_r+1}{2} + \frac{\epsilon_r-1}{2}\left[1+12\frac{h}{W}\right]^{-\frac{1}{2}} \tag{2}$$

$$\frac{\Delta L}{h} = 0.412\frac{\left(\epsilon_{r_{eff}}+0.3\right)\left(\frac{W}{h}+.264\right)}{\left(\epsilon_{r_{eff}}-.258\right)\left(\frac{W}{h}+0.8\right)} \tag{3}$$

$$L = \frac{c}{2f_r\sqrt{\epsilon_{r_{eff}}}} - 2\Delta L \tag{4}$$

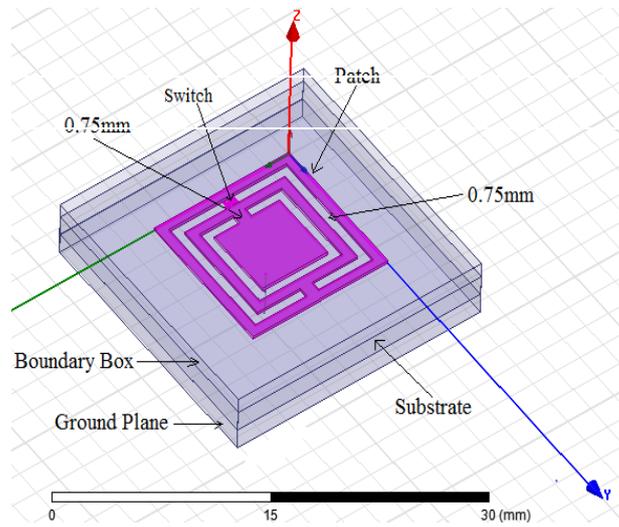

Fig. 3. Top view of proposed Antenna.

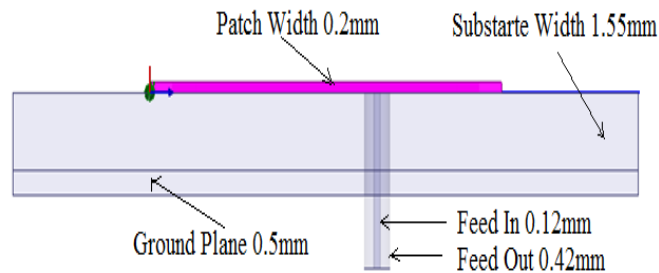

Fig.4 Side view of proposed Antenna.

Thus a convenient patch is designed using above mentioned equations. In this paper we have taken square patch i.e. the length and width will be same and is taken as 11.60mm. Thus a square patch is of dimension $11.60 \times 11.60 mm^2$ which is shown in the above figure 3: . Here a Split Ring Resonator (SSR) is taken in the patch to increase the gain. The taken out SSR along with its dimension is shown in the figure 3. . Respectively the top view and side view of the design is also shown in the above figures. Table 1 describes the material of the patch which is copper with permittivity $\varepsilon_r = 1$ and that of the substrate is of FR4 epoxy with permittivity $\varepsilon_r = 4.4$.

### III. SIMULATION RESULT AND DISCUSSION

We used High Frequency Structure Simulator (HFSS) for simulation of our design which is very good simulator for RF antennas. After simulating the designed antenna, the results which we obtained is displayed in figure (5-8).

Table-1 value of Patch and Substrate for Proposed Antenna.

| Material | |
|---|---|
| Patch | Copper with $\varepsilon_r = 1$ |
| Substrate | FR4 epoxy with $\varepsilon_r = 4.4$ |

Table-2 Design Parameter for Proposed Antenna.

| Parameter | Values (mm) |
|---|---|
| Width of patch (w) | 11.60 |
| Length of patch (L) | 11.60 |
| Height of patch (t) | 0.2 |
| Width of substrate (ws) | 20.6 |
| Length of substrate (Ls) | 20.6 |
| Height of substrate (Hs) | 1.55 |
| Dielectric constant ($\varepsilon_r$) | 4.4 |
| Effective dielectric constant ($\varepsilon_{rff}$) | 3.75 |
| Height of ground (Hg) | 0.5 |
| Height of patch (hp) | 0.2 |
| Extended patch length (ΔL) | 0.692 |
| Inner radius of co-axial feed | 0.12 |
| Outer radius of co-axial feed | 0.42 |

In table-2 show the precise dimension of various parameter of proposed design of antenna. Table 3 gives the complete description of the various parameters of the proposed antenna design such as $S_{11}$, VSWR, gain and bandwidth for different obtained bands.

Table-3 Simulation Result for Proposed Antenna.

| Band | Parameter | Switch OFF | Switch ON |
|---|---|---|---|
| 5.9 GHz | $S_{11}$ | -15.16 dB | -13.82 dB |
|  | VSWR | 1.65 | 1.51 |
|  | Gain | 0.811 dB | 1.211 dB |
|  | Bandwidth | 210 MHz | 230 MHz |
| 5.0 GHz | $S_{11}$ | - | -26.60 |
|  | VSWR | - | 1.099 |
|  | Gain | - | 1.211 dB |
|  | Bandwidth | - | 127 MHz |
| 8.1 GHz | $S_{11}$ | -33.05 dB | - |
|  | VSWR | 1.055 | - |
|  | Gain | 0.39 | - |
|  | Bandwidth | 190 MHz | - |

Fig. 5 &6 shows the total gain of antenna for two states of switch. One when the switch is OFF the total gain of antenna is 0.811 dB and other when the switch is ON the total gain of antenna is 1.211 dB.

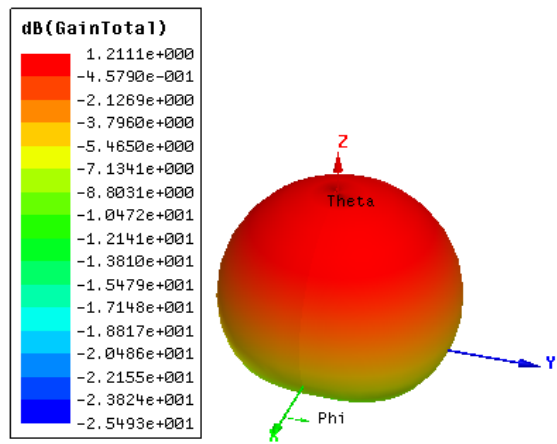

Fig.5 Total Gain when switch is ON.

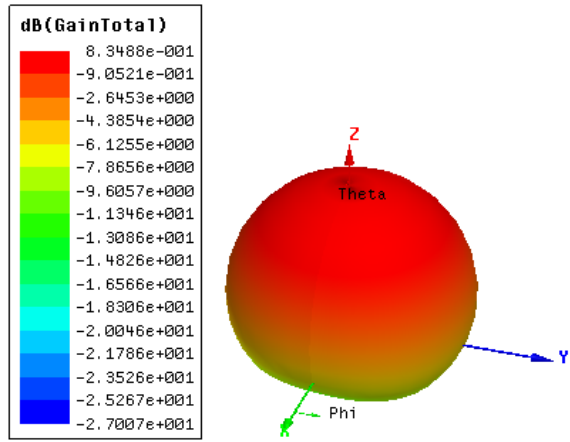

Fig.6 Total Gain when switch is OFF.

Figure 7 & 8 is the comparison plot of $S_{11}$ parameter and VSWR respectively for switch OFF and switch ON and the values of those plots are given in table 9.

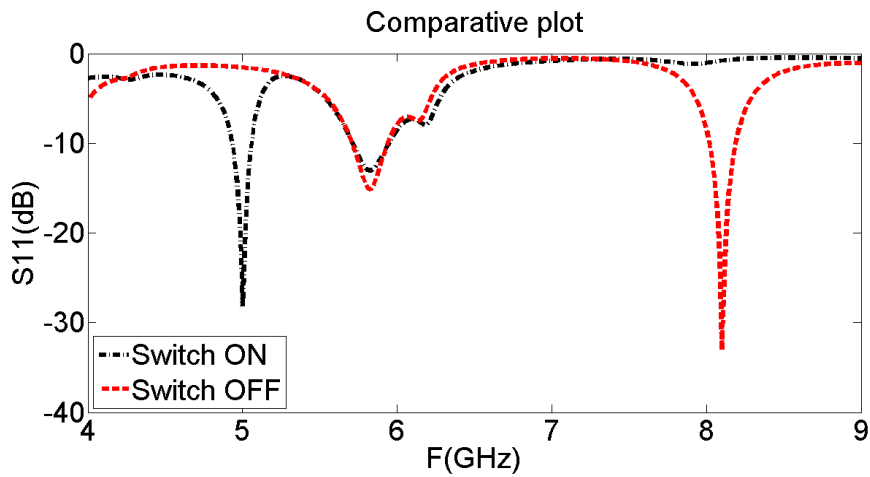

Fig.7 Comparison of S11 Parameter when switch is OFF and switch is ON.

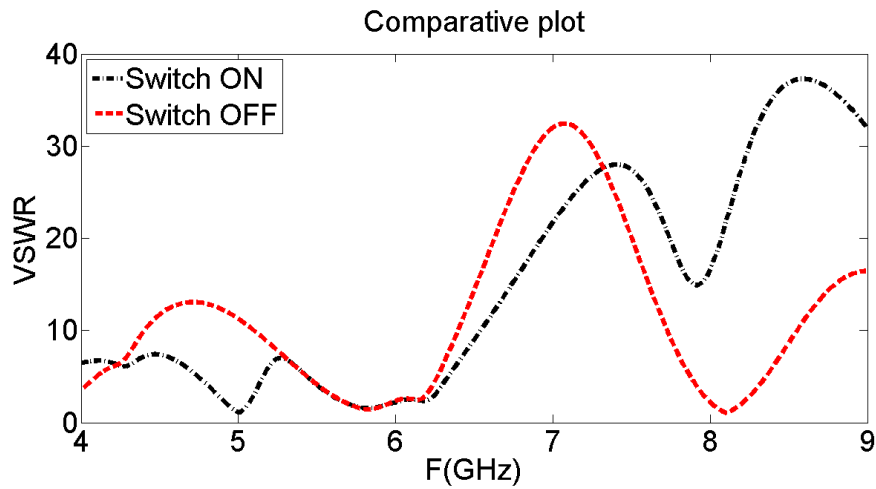

Fig.8 Comparison of VSWR when switch is OFF and switch is ON.

Table-4 Resultant output when switch is OFF.

| Resultant Output when switch is OFF | | |
|---|---|---|
| Parameter | 1st | 2nd |
| Frequency in GHz | 8.1 | 5.9 |
| Minimum return loss (S11) in dB | -33.05 | -15.60 |
| VSWR | 1.055 | 1.65 |

Table-5 Resultant output when switch is ON.

| Resultant Output when switch is ON | | |
|---|---|---|
| Parameter | 1st | 2nd |
| Frequency in GHz | 5.0 | 5.9 |
| Minimum return loss (S11) in dB | -26.60 | -13.82 |
| VSWR | 1.099 | 1.51 |

## IV. CONCLUSION

Antennas have become a rapidly growing area of research in the recent wireless communication system .The reason behind is that the compact size, less weight, low cost, low profile, easy production and easy addition of feed Network. Here Split ring resonator type microstrip patch is designed for multiband applications. Here the simulation is carried out at centre frequency 6GHz.Output result contains two bands out of which one band is at 5.9GHz and other one is at 8.1MHz so the antenna can used for C Band and X Band Applications. Further design can be modified to have a multiband for other applications in Ku Band, K Band, Ka Band. The results obtained for reconfigurable microstrip patch antenna are more efficient as compared to other antennas so far designed for similar applications. In above design we have used FR4 as a substrate. Instead of FR4 as substrate, the artificial material, meta material along with multilayer substrate (super substrate) can be used further to improve results
.